\documentclass[twocolumn,aps,amsmath,amssymb,nofootinbib,pre]{revtex4}
\usepackage[normalem]{ulem}
\newcommand{\stk}[1]{\ifmmode\text{\sout{\ensuremath{#1}}}\else\sout{#1}\fi}
\usepackage{amssymb,amsfonts,textcomp}
\usepackage{array}
\usepackage{supertabular}
\usepackage{hhline}
\usepackage{graphicx}
\usepackage{xcolor}
\usepackage{soul}

\begin{document}

\title{Synchronization transition of the second-order Kuramoto model 
on lattices}

\author{G\'eza \'Odor and Shengfeng Deng}
\address{Centre for Energy Research, Institute of Technical Physics
and Materials Science, \\ P. O. Box 49, H-1525 Budapest, Hungary}

\begin{abstract}

The second-order Kuramoto equation describes synchronization
of coupled oscillators with inertia, which occur in power grids
for example. Contrary to the first-order Kuramoto equation it's
synchronization transition behavior is much less known. In case
of Gaussian self-frequencies it is discontinuous, in contrast to 
the continuous transition for the first-order Kuramoto equation.
Here we investigate this transition on large 2d and 3d
lattices and provide numerical evidence of hybrid phase transitions, 
that the oscillator phases $\theta_i$, exhibit a crossover, while 
the frequency spread a real phase transition in 3d. 
Thus a lower critical dimension $d_l^O=2$ is expected for the
frequencies and $d_l^R=4$ for the phases like in the massless case.
We provide numerical estimates for the critical exponents, 
finding that the frequency spread decays as $\sim t^{-d/2}$ 
in case of aligned initial state of the phases in agreement with
the linear approximation.
However in 3d, in the case of initially random distribution of 
$\theta_i$, we find a faster decay, characterized by 
$\sim t^{-1.8(1)}$ as the consequence of enhanced nonlinearities 
which appear by the random phase fluctuations.
\end{abstract}

\maketitle

\section{Introduction}

Synchronization within interacting systems is an ubiquitous phenomenon 
in nature. It has been observed in biological, chemical, physical, and 
sociological systems. Much effort has been dedicated for theoretical 
understanding of its general features \cite{Pik,Acebron,ARENAS200893}.
A paradigmatic model of $N$ globally coupled oscillators was introduced 
and solved in the stationary state in the limit $N\to\infty$ by 
Kuramoto~\cite{kuramoto} and later the macroscopic evolution of the system 
was shown to be governed by a finite set of nonlinear ordinary differential 
equations~\cite{oa}.
An interesting property of the so-called first-order Kuramoto model is that 
it has a continuous phase transition, with a diverging correlation size, 
separating a synchronized phase from an unsynchronized one.
Due to the chaoticity, emerging from nonlinearity, it obeys a scaling theory, 
analogous to stochastic systems at the critical point and the whole set 
of critical exponents are known~\cite{kuramoto,oa,chate_prl,cmk2016}. 
The corresponding universality class is termed as mean-field since, due to 
the all-to-all coupling, the individual oscillators interact with a mean-field 
of the rest of the oscillators.
A challenging research direction aims at studying the possibility and 
nature of synchronization transitions in extended systems, 
where oscillators are fixed at regular lattice sites of finite dimension $d$ 
and the interaction, in the extreme case, is restricted to 
nearest-neighbors~\cite{sakaguchi,HPCE,chate_prl,Kurcikk}.

The so-called second-order Kuramoto model was proposed to describe
power grids, analogous to the swing equation of AC circuits~\cite{fila}.
This is the generalization of the Kuramoto model \cite{kuramoto} with 
inertia. One of the main consequences of this inertia is that
the second-order phase synchronization transition, observed in
the mean-field models of the massless first-order Kuramoto models 
turns into a first-order one~\cite{TL97}.

However, in lower dimensions this has not been studied systematically.
In~\cite{POWcikk} numerical integration on 2d lattices suggested 
crossover transitions, with hysteresis in case of the phase 
order parameter. 
Note, that due to the inherent heterogeneity of the quenched 
self-frequencies $\omega_i(0)$ of the nodes, rare-region effects
may occur, leading to frustrated synchronization and chimera
states~\cite{POWcikk,Frus,FrusB,USAEUpowcikk}.

As real power grids are connected via complex networks, topological
heterogeneity are also present, which can smear a phase transition, 
strengthening possible rare-region effects. 
However, even if topological heterogeneity are not present, it 
is still not proven yet whether the massive model exhibit real phase 
transitions at low dimensions.
Only conjectures, that the massive model has the same lower and upper 
critical dimensions as the first-order
Kuramoto model \footnote{in case of single peaked self-frequency 
distribution}, are available.
Accordingly, mean-field phase transition for $d \ge 4$ of
the phase order parameter and a crossover below it~\cite{Acebron} 
should occur \footnote{However, even the $d_c=4$ conjecture is debated, 
some studies concluded $d_c=5$ or higher ~\cite{HPCE}.}.
Thus, the upper and lower critical dimensions may be identical: 
$d_c = d_l = 4$.

For the frequency entrainment of the massless model the lower 
critical dimension is expected to be at $d_l^O=2$ similar to
the Mermin-Wagner theorem~\cite{PhysRevLett.17.1133} for the 
planar XY spin model and is supported by finite size scaling 
analysis~\cite{HPCE}.
Thus, for intermediate dimensions $2 < d <4$, real, nontrivial 
continuous phase transition should occur.
Analogously, for the massive 
case~\cite{POWcikk,Powfailcikk,USAEUpowcikk}, entrainment phase 
transition is also expected for dimensions $2 < d <4$, as a 
very recent power-grid study~\cite{USAEUpowcikk} has indicated it 
for networks with graphs dimensions $2 < d < 3$.

This has recently been published for the high voltage 
power-grid networks of the USA and Europe and now we shall
investigate it in case of pure 2d and 3d lattices, using 
finite size scaling.
In that work the linear approximation, which is expected to be valid
for large couplings, provided a frequency spread decay law 
$\Omega \sim t^{-d/2}$~\cite{USAEUpowcikk}. 
Now we test the applicability of this approximation at the phase 
transition points.

Besides the dynamical scaling the frequency order
parameter exhibited a hysteresis and a discontinuity~\cite{USAEUpowcikk},
which is known in statistical physics~\cite{Cardy_1981},
termed as hybrid or mixed type of phase transition,
for example at tricriticality~\cite{PhysRevE.70.026114,Gras-av},
or in other nonequilibrium 
systems~\cite{PhysRevLett.106.128701,PhysRevE.87.032106,odorSim21}.
Now we investigate in detail this transition, which arises by
the inertia in the Kuramoto model and results in hysteresis as
we change the synchronization level of the initial states.

\section{Models and methods}

\subsection{The second-order Kuramoto model}

Time evolution of power grids synchronization is described by the swing 
equations~\cite{swing}, set up for mechanical elements with inertia.
It is formally equivalent to the second-order Kuramoto equation~\cite{fila},
for a network of $N$ oscillators with phases $\theta_i(t)$:
\begin{eqnarray}\label{kur2eq}
\dot{\theta_i}(t) & = & \omega_i(t) \\
\dot{\omega_i}(t) & = & \omega_i(0) - \alpha \dot{\theta_i}(t) 
+ K \sum_{j=1}^{N} A_{ij} \sin[ \theta_j(t)- \theta_i(t)] \ . \nonumber
\end{eqnarray}
Here $\alpha$ is the damping parameter, which describes the power dissipation,
or an instantaneous feedback~\cite{Powfailcikk}, $K$ is the global coupling, 
related to the maximum transmitted power between nodes and $A_{ij}$, which 
is the adjacency matrix of the network, containing admittance elements. 
The quenched self-frequency of the $i$-th oscillator is $\omega_i(0)$, 
which describes the power in/out of a given node when Eq.~(\ref{kur2eq}) 
is considered to be the swing equation of a coupled AC circuit, but
here we have chosen it zero centered Gaussian random variable as
rescaling invariance of the equation allows to transform it within 
a rotating frame.

In our present study the following parameter settings were used:
the dissipation factor $\alpha$, is chosen to be equal to $0.4$ to 
meet expectations for power-grids, with the $[1/s]$ inverse time 
physical dimension assumption.
For modeling instantaneous feedback, or increased damping 
parameter we also investigated the $\alpha = 3.0\, [1/s]$ case, 
similarly as in~\cite{Powfailcikk,USAEUpowcikk}.

To solve the differential equations in general we used the adaptive 
Bulirsch‐Stoer stepper~\cite{boostOdeInt}, which provides more precise 
results for large $K$ coupling values than the Runge-Kutta method.
The solutions depend on the $\omega_i(0)$ values and become chaotic,
especially at the synchronization transition, and thus to obtain 
reasonable 
statistics, we needed strong computing resources, using parallel codes 
running on GPU clusters. The corresponding CUDA code allowed us to 
achieve $\sim 100\times$ speedup on GeForce GTX 1080 cards as compared
to Intel(R) Core(TM) i7-4930K CPU @ 3.40GHz cores.
The details of the GPU implementation will be discussed in a separate 
publication \cite{texpaper}.

We obtain larger synchronization if the initial state is set 
to be fully synchronized, with phases: 
$\theta_i(0)=0$, but due to the hysteresis one can also investigate 
other uniform random distributions like: $\theta_i(0) \in (0,2\pi)$. 
The initial frequencies were set as: $\dot{\theta_i}(0)=\omega_i(0)$.

To characterize the phase transition properties, both the phase
order parameter $R(t)$ and the frequency spread $\Omega(t)$, termed the frequency order
parameter, will be studied. We measured the Kuramoto phase order parameter:
\begin{equation}\label{ordp}
z(t_k) = r(t_k) \exp{i \theta(t_k)} = 1 / N \sum_j \exp{[i \theta_j(t_k)}] \ ,
\end{equation}
by increasing the sampling time steps exponentially:
\begin{equation}
t_k = 1 + 1.08^{k} \ ,
\end{equation}
where $0 \le r(t_k) \le 1$ gauges the overall coherence and $\theta(t_k)$ is
the average phase. The set of equations (\ref{kur2eq}) were solved 
numerically for $10^3 - 10^4$ independent initial conditions, 
initialized by different $\omega_i(0)$-s and different
$\theta_{i}(0)$-s if a disordered initial phases were invoked. 
Then sample averages for the phases 
\begin{equation}\label{KOP}
R(t_k) = \langle r(t_k)\rangle
\end{equation} 
and for the variance of the frequencies
\begin{equation}\label{FOP}
\Omega(t_k,N) = \frac{1}{N} \sum_{j=1}^N (\overline\omega-\omega_j)^2 
\end{equation}
were determined, where $\overline\omega(t_k)$ denotes the mean
frequency within each respective sample.

In the steady state, which we determined by visual inspection of the mean 
values $R(t_k)$, we measured the standard deviations $\sigma(R)$ of the order 
parameters $R(t_k)$ in order to locate the transition point by 
fluctuation maxima. While the transition point for $\Omega(t_k,N)$ is 
characterized by a sudden drop of the $\Omega(t\to\infty,N)$ or by an 
emergence of an algebraic decay of $\Omega(t)$ as we increase $K$. 
In case of the first-order Kuramoto equation the fluctuations of
both order parameters show a maximum at the respective transition 
points~\cite{Flycikk}.
For the second-order Kuramoto, only the $\sigma(R(t_k))$ seems to have
a peak at $K'_c$, while for $\Omega(t_k,N)$ we located a different 
transition point $K_c$, where the saturation to steady state value 
changed to a decay in the $t\to\infty$ limit.

\subsection{Linear approximation for the the frequency entrainment}

In Ref.~\cite{USAEUpowcikk}, we showed that, similar to the first-order 
Kuramoto model, 
the frequency order parameter \eqref{FOP} decays as 
$\Omega \propto t^{-d/2}$  on a $d$-dimensional lattice in the large system size 
and large coupling constant limit \cite{HPCE}. By applying the linear approximation 
$\sin(x) \propto x$ and casting the continuum second-order Kuramoto equations into 
the momentum space, the phase velocity [$\omega(\mathbf{x},t)\equiv\dot{\theta}
(\mathbf{x},t)$] is obtained \cite{USAEUpowcikk}
\begin{align}
	\omega(\mathbf{k},t)=&e^{-\frac{1}{2} t (\alpha +\Delta )} 
	\Big[\omega(\mathbf{k},0) \big(( \Delta +2 -\alpha) e^{\Delta  t}
	\nonumber\\
	& +\alpha +\Delta -2\big)-2 K k^2 
	\theta(\mathbf{k},0) \left(e^{\Delta  t} -1\right)\Big]/2\Delta\,,
	\label{velocity}
\end{align}
where $\Delta=\sqrt{\alpha^2-4K k^2}$. When initial disordered condition is 
considered, say $\theta(\mathbf{x},0)$ is uniformly distributed over 
$(0,\theta_\mathrm{max})$, one has $\langle \theta(\mathbf{x}) 
\theta(\mathbf{x}')\rangle = \theta_\mathrm{max}^2/4$, suggesting that
$\langle \theta(\mathbf{k},0) \theta(\mathbf{k}',0)\rangle=\mathbf{\delta}^d(\mathbf{k})
\mathbf{\delta}^d(\mathbf{k}')$. Hence, in the linear approximation, disorder
from the initial condition doesn't affect the frequency spread (note that
$\langle \omega(\mathbf{k},0) \theta(\mathbf{k},0) \rangle =0$) and we 
have (the same as in Ref.~\cite{USAEUpowcikk}):
\begin{align}
\Omega(t)=&\frac{1}{L^d}\int d^d\mathbf{x} \langle 
\left[\omega(\mathbf{x},t)-\bar{\omega}(t)\right]^2\rangle 
	   \nonumber\\
	 =&C_d\int_{2\pi/L}^{\pi/a}dk k^{d-1}\frac{e^{-t (\alpha 
	 +\Delta )}}{4 \Delta ^2}
	 \Big[\alpha +\Delta -2 \nonumber\\
	 &+(\Delta-\alpha +2) e^{\Delta  t}\Big]^2\,,
	\label{spreadlin}
\end{align}
where $\bar{\omega}(t)$ denotes the spatial average of 
$\omega(\mathbf{x},t)$, while $a$ and $C_d$ are the lattice spacing and 
the geometric factor, respectively.

As shown in Ref.~\cite{USAEUpowcikk}, Eq.~\eqref{spreadlin} gives rise to the
$t^{-d/2}$ law that manifests a rapid cutoff for large couplings in a
typical finite system, whereas in the regime where a linear
approximation is invalid, weak couplings fail to maintain a narrow
frequency entrainment and $\Omega$ is bound to be stationary after some
time. Hence, a frequency entrainment phase transition, from finite
stationary $\Omega$ value to infinitely decaying $\Omega$ is expected.

\section{Synchronization transition in 2d}

We solved the system of equations (\ref{kur2eq}) on large square 
lattices with periodic boundary conditions for linear sizes:
$L= 200,\, 400,\, 1000,\, 2000$. The self-frequencies were chosen 
randomly from a zero centered Gaussian distribution with unit variance.
The order parameters were calculated by ensemble averages over many
samples.

\subsection{Frequency entrainment phase transition}

It is known that the frequency order parameter (\ref{FOP}) decays as 
$\Omega \propto t^{-d/2}$ in case of the first-order Kuramoto
model in the large coupling limit if we start from a random initial state
\cite{HPCE}. We have also shown that the same is true
for the second-order Kuramoto model in the linear approximation in
\cite{USAEUpowcikk}. Now we investigate this at the neighborhood of 
frequency entrainment transition point. 

\begin{figure}[!htbp]
	\centering
	\includegraphics[width=0.49\textwidth]{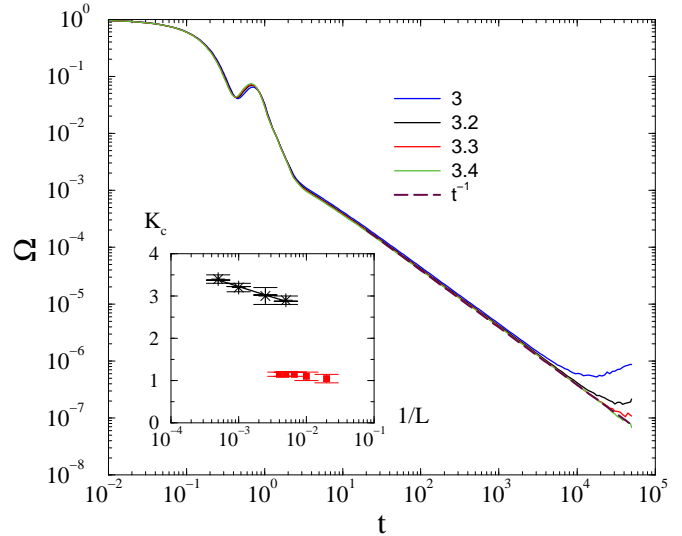}
	\caption{The frequency spread in 2d at $\alpha=3$ for different
	$K$ values, shown by the legends, for $L=2000$, in case of 
        ordered initial conditions. 
        The dashed line marks a numerical fit at the critical point
        $K_c=3.4(1)$ 
        with $t^{-d/2}$. Inset: finite size scaling of the frequency 
        entrainment transition point $K_c$ for various system sizes in 
	2d (black asterisks) and 3d (red boxes), for $\alpha=3$ and 
	ordered initial conditions.
        One can see a logarithmic growth in 2d and a convergence to
        $K_c=1.15(5)$ constant value in 3d. }
	\label{O2D_L2000_a3}
\end{figure}

As Fig.~\ref{O2D_L2000_a3} shows the density decays as $\Omega\propto t^{-1}$ 
at the critical coupling strength: $K_c=3.4(1)$ in the case of ordered phase
initial conditions, for $\alpha=3$ damping factor.
The decay behavior follows the same power law for $K \ge K_c$
before the finite size cutoff can take effect, and we see a
saturation to finite values for $K < K_c$.

\begin{figure}[!htbp]
        \centering
        \includegraphics[width=0.49\textwidth]{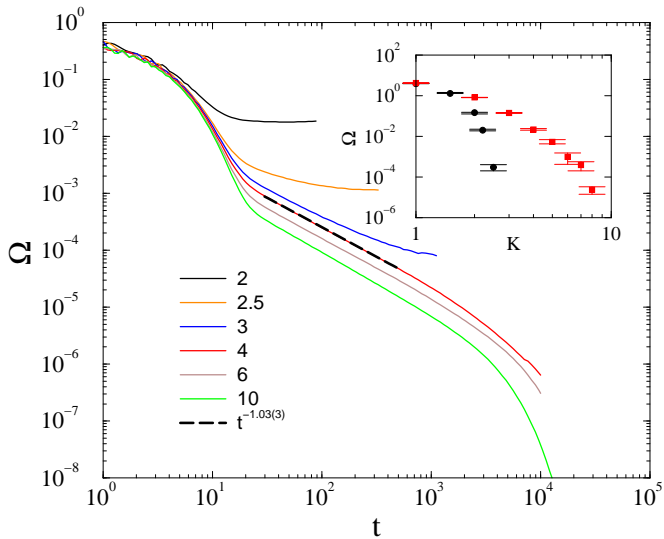}
        \caption{The frequency spread in 2d at $\alpha=0.4$ for different
        $K$ values, shown by the legends, for $L=2000$, using ordered 
        initial conditions.
        The dashed line marks a numerical fit at the critical point 
        $K_c = 3.5(5)$ with $t^{-1.03(3)}$.
	Inset: Steady state values obtained by starting from ordered
        (black bullets) and disordered (red boxes) initial conditions.
	}
        \label{O2D_L2000_a0.4}
\end{figure}

The same is true for $\alpha=0.4$: following a longer initial transient
we can see a decay at $K_c = 3.5(5)$ characterized by $\Omega\propto t^{-1}$.
as shown by  Fig.~\ref{O2D_L2000_a0.4}. An exponential finite size cutoff
occurs already for $t>1000$ in contrast to the $\alpha=3$ case, where
this happened above $t > 10^4$.

For smaller system sizes the $K_c$-s do not move a lot, as we can
see from the inset of Fig.~\ref{O2D_L2000_a3}. 
The available data precision restricts finite size scaling, but 
still we attempted it as shown in the inset of Fig.~\ref{O2D_L2000_a3}. 
Assuming a logarithmic growth dependence, which is expected at the
lower critical dimension~\cite{HPCE} we obtained
$K_c(1/L) \propto -1.7(1)\ln(1/L)$. 

In the case of \textit{fully disordered} initial conditions, 
$\theta_i(t=0)\in (0,2\pi)$, we found the same behavior as in case of 
the fully phase synchronized starts, as one can see on 
Fig.~\ref{O2D_L2000_a3_TM2} for $\alpha=3$ and
Fig.~\ref{O2D_L2000_a0.4_TM2} for  $\alpha=0.4$
shown in the Appendix.

The steady state values, appearing for $t > 10^4$ near the 
critical point of the $\alpha=0.4$ damping factor case are also 
determined and plotted in the inset of Fig.~\ref{O2D_L2000_a0.4} for
$L=200$. We can see two branches, depending on the initial
conditions. The upper branch corresponds to the disordered,
the lower to the fully ordered initial states. Thus we can 
see a hysteresis like behavior near the phase transition.
But the approach of $\Omega(K \to K_c)$ is rather smooth,
which is not surprising at a crossover point.

\begin{figure}[!htbp]
        \centering
        \includegraphics[width=0.49\textwidth]{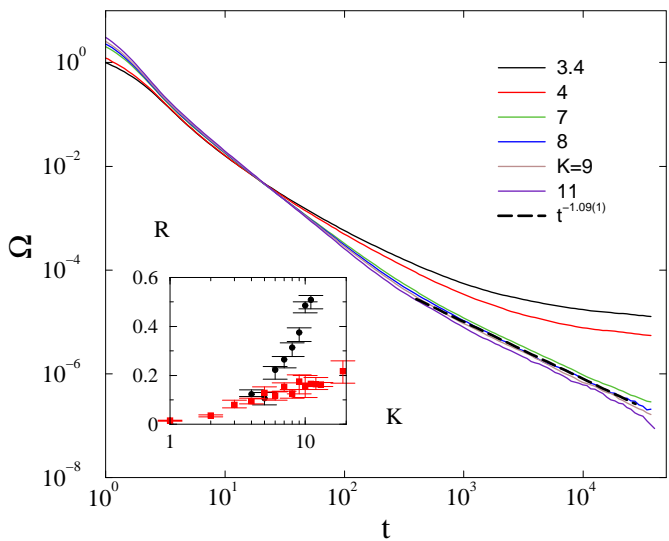}
        \caption{The frequency spread in 2d at $\alpha=3$ for different
        $K$ values, shown by the legends, for $L=2000$, in case of 
        disordered initial conditions.
        The dashed line marks a numerical fit at the critical point 
        at $K_c = 8.0(5)$ with $t^{-1.09(5)}$.
        Inset: Part of the hysteresis loop of $R$ in 2d obtained 
	by ordered (black bullets) and disordered (red boxes) initial 
        conditions for $\alpha=3$ and $L=200$.}
        \label{O2D_L2000_a3_TM2}
\end{figure}

\subsection{Phase order parameter transition}\label{sect:2D:R}

\begin{figure}[!htbp]
        \centering
        \includegraphics[width=0.49\textwidth]{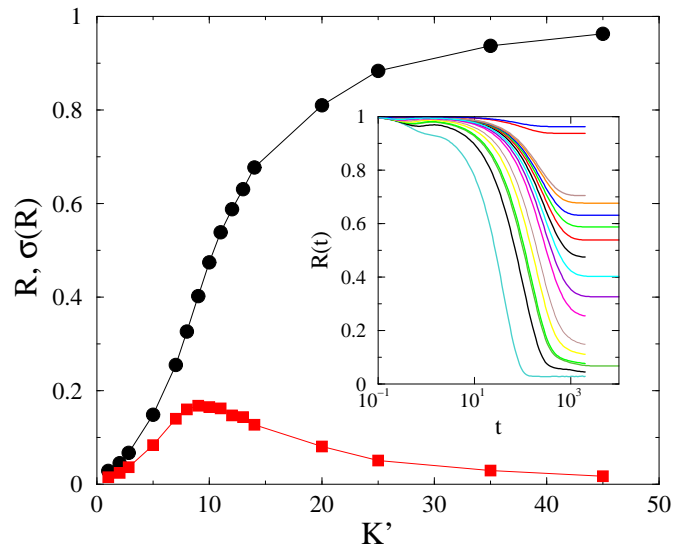}
        \caption{Steady state Kuramoto order parameter in 2d and
	      its variance at $\alpha=3$ at different $K$ values 
              for $L=200$. Inset: $R(t,L=200)$ for $K = $ 1, 2, 3, 5, 7, 
              8, 9, 10, 11, 12, 13, 14, 20, 25, 35, 45 (bottom to top curves). }
        \label{beta2D_L200_a3}
\end{figure}

We determined the steady state values of $R(t,L)$ by starting the
systems from fully phase coherent states up to
$t_{max}=10^4\mbox{--}5\times
10^4$ followed by a visual inspection. For a certain system size
$L$, we obtain the dependence of the stationary phase order parameter 
$R_\infty$ on $K$. Fig.~\ref{beta2D_L200_a3} shows one such example for 
$L=200$ and $\alpha=3$ in 2d.  The transition point $K'_c$ then could either 
be located by the peaks of $\sigma(R)$ as chaoticity take maximum value 
at $K'_c$~\cite{USAEUpowcikk,KKI18deco,Flycikk}, or be estimated by the
half value $R(L,K'_c)\simeq 0.5$.
However, this transition point did not coincide with the critical
point $K_c$ determined by the order parameter $\Omega$.

\begin{figure}[!htbp]
	\centering
	\includegraphics[width=0.49\textwidth]{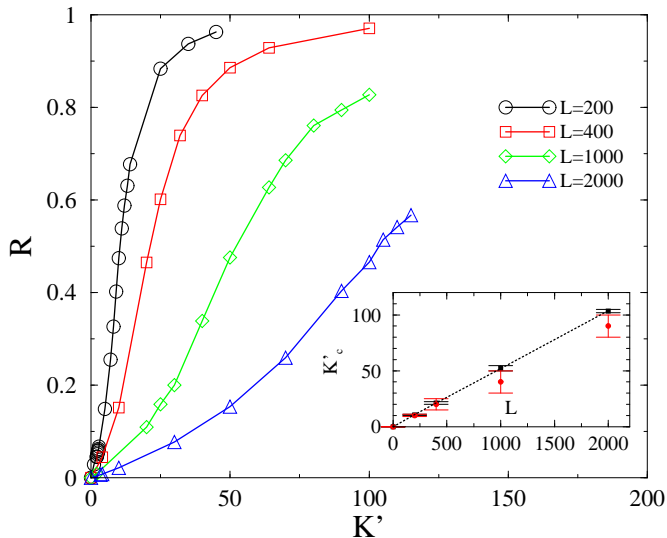}
	\caption{Finite-size behavior of $R$ in 2d for $\alpha=3$ and ordered 
        initial conditions, shows a crossover. 
	Inset: finite-size scaling of $K'_c$ as estimated by the half
	values of $R$ (black boxes) and by the $\sigma(R)$ peaks 
        (red bullets) exhibit a linear growth.}
	\label{crossover2d}
\end{figure}

As remarked in the Introduction section, we conjecture that the Kuramoto
order parameter $R$ exhibits a real discontinuous transition above
$d_l^{R} > 4$, while for $d \le d_l^{R}$ a crossover transition ensues. 
To verify this conjecture, we estimate the transition point $K'_c$ and 
check if it diverges in an infinite system. 
The crossover transition nature (rather than a real transition) is 
immediately clear as demonstrated by Fig.~\ref{crossover2d}, in 
which we see an evident shift of the transition point as the system 
size is varied. The $\sigma(R)$ also become wider and wider as we
increase the size.

Particularly, the inset suggest that the transition point shifts
linearly with $L$ in 2d [$K'_c(L)\propto L$]. 
Hence the transition points exhibit a power-law growth 
with exponents, suggesting that $K'_c(L)\to 0$ as
$L\to 0$ and $K'_c(L) \to \infty$ as $L\to \infty$.

For disordered initial conditions we can find much lower steady state
values indicated by the inset of Fig.\ref{O2D_L2000_a3_TM2}. 
The hysteresis loop closes at very large $K$ values only, as 
was also demonstrated in~\cite{USAEUpowcikk} for power-grid networks.

\section{Synchronization transition in 3d}

In 3d, following the results of the first-order Kuramoto model
we expect a real phase transition of the frequency order parameter,
but a crossover for the phases.
Similarly to 2d we solved the system of equations (\ref{kur2eq}) 
on large cubic lattices with periodic boundary conditions for 
linear sizes: $L=50,\, 100,\, 150,\, 200,\, 250$ in order to perform
finite size analysis.

\subsection{Frequency entrainment phase transition}

\begin{figure}[!htbp]
        \centering
        \includegraphics[width=0.49\textwidth]{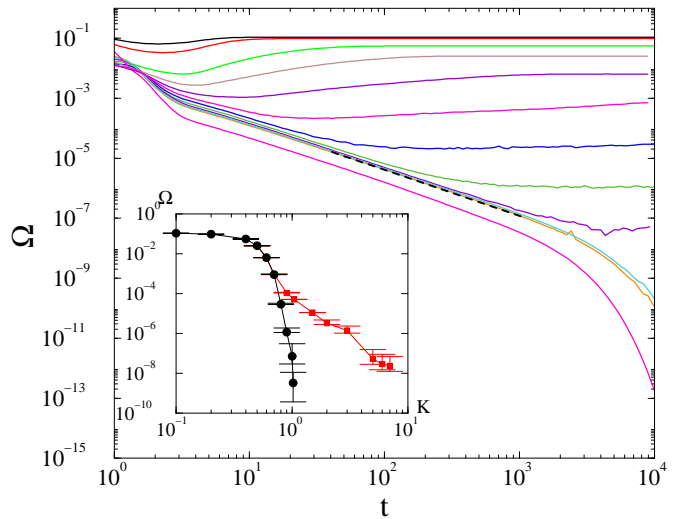}
        \caption{The frequency spread in 3d at $\alpha=3$ for
        $K=$ 0.1, 0.2, 0.4, 0.5, 0.6, 0.7, 0.8, 0.9, 1, 1.05, 1.1, 2 
        (top to bottom curves)
        for $L=200$ linear sized lattices and phase ordered initial 
        conditions.
        The dashed line marks a numerical fit at the critical point 
        $K_c = 1.02(2)$ with $t^{-d/2}$.
        Inset: Steady state values obtained by starting from ordered 
        (black bullets) and disordered (red boxes) initial conditions. }
        \label{O3D_L200_a3_TM0}
\end{figure}

In case of \textit{phase ordered initial states} the frequency spread decays
with the law $\Omega(t)\propto t^{-d/2}$ above $K_c\simeq 1.1$, 
followed  by a finite-size cutoff as shown on Fig.~\ref{O3D_L200_a3_TM0}
for $L=200$. Doing the finite-size scaling of the transition point,
we find that $K_c$ does not change within error margins for $L \ge 150$ 
and we estimate a finite value: $K_c = 1.15(5)$ as shown in the inset of
Fig.~\ref{O2D_L2000_a3}. 

\begin{figure}[!htbp]
        \centering
        \includegraphics[width=0.49\textwidth]{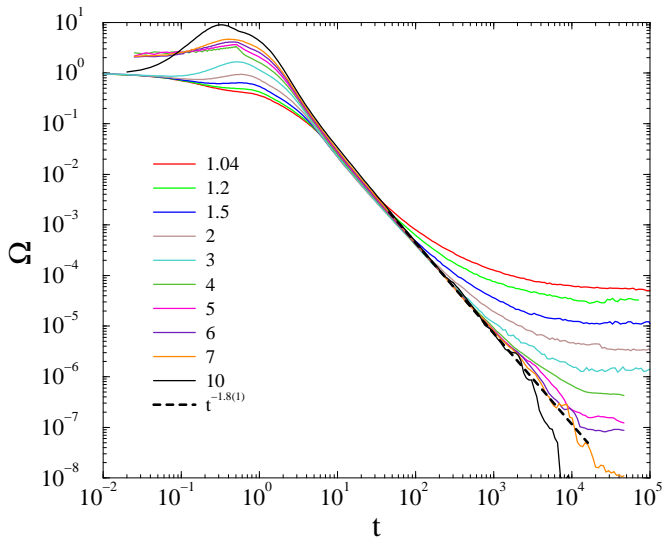}
        \caption{The frequency spread in 3d at $\alpha=3$ for different
        $K$ values, shown by the legends, for $L=200$ and disordered
        initial conditions. The dashed line marks a numerical
        fit at the critical point $K = K_c \simeq 7$ with $t^{-1.8(1)}$.
        }
        \label{O3D_L200_a3_TM2}
\end{figure}

However, in case of the \textit{fully random phase initial condition}
the decay at the critical point seems to deviate from the $t^{-d/2}$ law.
It can be be fitted by $\Omega(t)\propto t^{-1.8(1)}$ at
$K = K_c \simeq 7$ as shown on Fig.~\ref{O3D_L200_a3_TM2}.
Note, that around criticality, in the $t > 10^3$ region, where
finite-size effects emerge, the slope of curves increases, suggesting
a nontrivial correction like in case of the first-order Kuramoto
model~\cite{Kurcikk}. Due to the limited computing power, this
excludes the possibility to see a crossover towards a 
$\Omega(t)\propto t^{-d/2}$  asymptotic behavior obtained by the 
linear approximation.
We have investigated this behavior for other levels of randomness 
in the initial state $\theta_{max} = 1,\, 1.75,\,1.9$, 
but found it only at the fully random phase case.

In the case with disordered initial conditions, the level-off of
$\Omega(t)$, thus $K_c$ also occurs at a much higher couplings, 
than in the ordered initialization case as the consequence of the 
phase transition. Therefore, we conjecture a possible
different scaling behavior, if any, at the higher $K_c$ value.
The steady state behavior of $\Omega$ is also shown in the inset of
Fig.~\ref{O3D_L200_a3_TM0}. At first sight it may not suggest 
a discontinuous transition, but as we applied log-log scales, to
observe the rapid changes two branches emerge and we can see the 
occurrence of a wide hysteresis loop as the consequence of 
different initial conditions.

\subsection{Phase order parameter transition}

\begin{figure}[!htbp]
	\centering
	\includegraphics[width=0.49\textwidth]{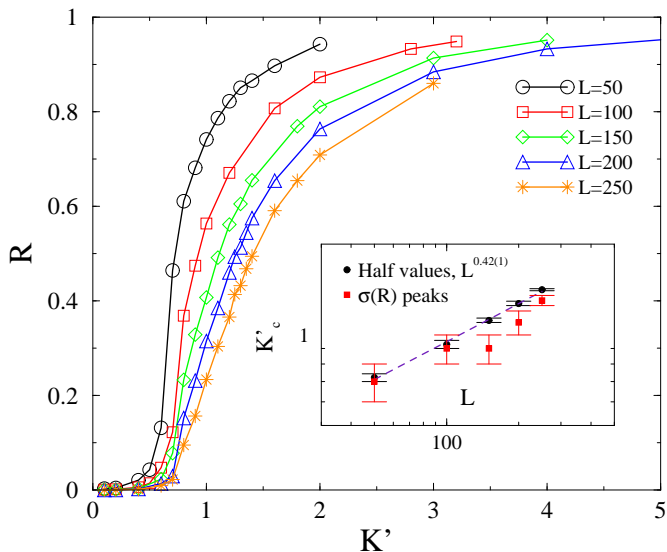}
	\caption{Finite-size behavior of $R$ in 3d for $\alpha=3$ and ordered 
        initial
	conditions, shows a crossover. Inset: finite-size scaling
        of $K'_c$ as estimated by the half values of $R$ (black bullets) 
        as well as by the $\sigma(R)$ peaks (red boxes)
        exhibit a power-law growth.}
	\label{crossover3d}
\end{figure}

We determined the Kuramoto order parameter values in the steady
state for cubes with linear sizes $L=50, 100, 150, 200, 150$, 
using $\alpha=3$ and ordered initial conditions.
We display the results for $L=100$ in the Appendix; see
Fig.~\ref{beta3D_L100_a3}.
We attempted a finite size scaling analysis as in 2d, as shown on
Fig.~\ref{crossover3d}. The $\sigma(R)$ distributions are getting
very smeared as $L\to\infty$, making it difficult to locate the
peaks. But still, a reasonable power-law fit could be obtained, 
in agreement with the half value method described in Sec.~\ref{sect:2D:R}:
$K'_c \propto L^{0.42(1)}$, as one can see in the inset of 
Fig.~\ref{O2D_L2000_a3}. Thus, we still find a crossover behavior 
in 3d, with a lower $K'_c$ growth exponent than in 2d, which is 
expected to decrease as we increase the dimension approaching
the lower critical dimension.

\section{Conclusions}

We have performed an extensive numerical study of the synchronization
transition of the second-order Kuramoto model in 2 and 3 dimensions.
We provided numerical evidence that while the phase order parameter 
exhibits crossover transition, which diverges with the system size in
a power-law manner, the frequency spread order parameter exhibits
real phase transition in 3d. 
In the latter case the finite size dependence of the critical point is 
negligible on the system sizes we investigated and the transition point 
for an infinite system, estimated through extrapolation is also very 
close to those measured in finite systems, 
except for a logarithmic correction in 2d.
The transition of both order parameters exhibit hysteresis behavior 
though, with the steady state values, which depend on the initial 
conditions. 

However, the variance of $R$, representing chaoticity over initial
self frequency choices, has a smeared peak around the crossover
point, with growing spread as we increase $L$. This makes the location
of the crossover point hard to determine, but we used an alternative  
method, using half values of $R$, consistent with the peak locations,
as a reliable way to obtain it. 
While the $K'_c(L)$ grows linearly with $L$ in $2d$, in $3d$
we found a nontrivial power-law dependence: 
$K'_c(L) \propto L^{0.42(1)}$.
   
For the $\Omega$ order parameter we did not find a peak at the critical
point, in contrast with the case of the massless Kuramoto model, 
in agreement with a first-order type phase transition behavior.
However, we found asymptotic power-law decay: $\Omega(t) \propto t^{-d/2}$
for $K \ge K_c$, which agrees with the linear approximation result.
This allowed us to perform a crude finite size scaling of $K_c$,
which exhibits a logarithmic growth of $K_c$ in 2d and a saturation in 3d. 
Thus, similarly to the massless Kuramoto~\cite{HPCE} we claim
$d_l^O=2$ for the lower critical dimension. 

We also found a deviation from the linear approximation law
in $d=3$ in case of disordered initial states:
$\Omega(t) \propto t^{-1.8}$. 
This behavior might be the consequence of a slow crossover in time or 
the nonlinearities due to the phase fluctuations on the upper branch of 
the frequency order hysteresis curve.
This behavior may be observable in real-power grid situations, as we found
in \cite{USAEUpowcikk}, in case of larger damping factors.
For $\alpha=0.4$ this anomalous power-law region is less extended, but this
is true for all PL-s we see: the damping factor elongates the scaling
regions in agreement with the rescaling invariance of the differential
equation, shown in~\cite{USAEUpowcikk}.

The coexistence of power-law dynamics of $\Omega$ and the hysteresis in the
steady sates thus classifies this as a hybrid or mixed type 
of phase transition, which would be interesting to study further.

\acknowledgments{

We thank R\'obert Juh\'asz for the useful comments and Jeffrey Kelling
for the GPU code upgrades.
Support from the advantaged ELKH grant and the Hungarian National Research,
Development and Innovation Office NKFIH (K128989) is acknowledged.
Most of the numerical work was done on KIFU supercomputers of Hungary.
}

\section*{Appendix}

In this appendix we show results in 2d for the $\Omega(t)$ decay
solution in case of disordered initial conditions at $\alpha=0.4$.

\begin{figure}[!htbp]
        \centering
        \includegraphics[width=0.49\textwidth]{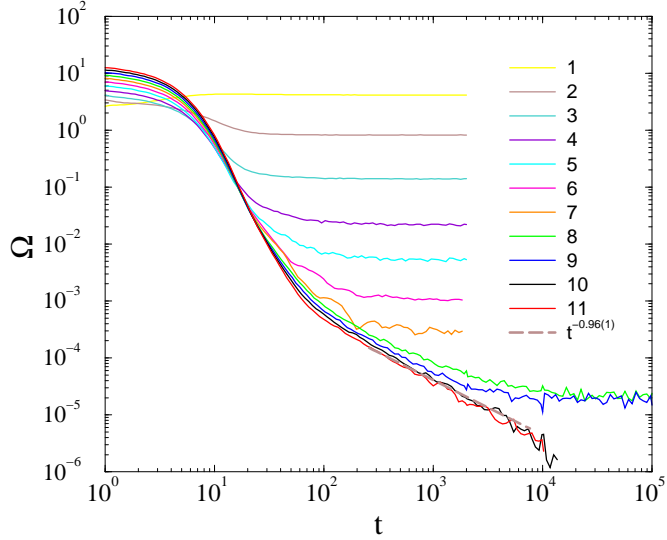}
        \caption{The frequency spread in 2d at $\alpha=0.4$ for different
        $K$ values, shown by the legends for $L=2000$, with disordered
        initial conditions.
        The dashed line marks a numerical fit at the critical point
        at $K_c = 9.5(5)$ with $t^{-0.96(5)}$.}
        \label{O2D_L2000_a0.4_TM2}
\end{figure}

Furthermore, we also plot the steady state behavior of $R$ in 3d, 
for $\alpha=3$, at $L=100$ and ordered initial conditions.
One can observe a peak in $\sigma(R)$ at $K\simeq 0.85$, where
$R \simeq 0.5$.

\begin{figure}[!htbp]
        \centering
        \includegraphics[width=0.49\textwidth]{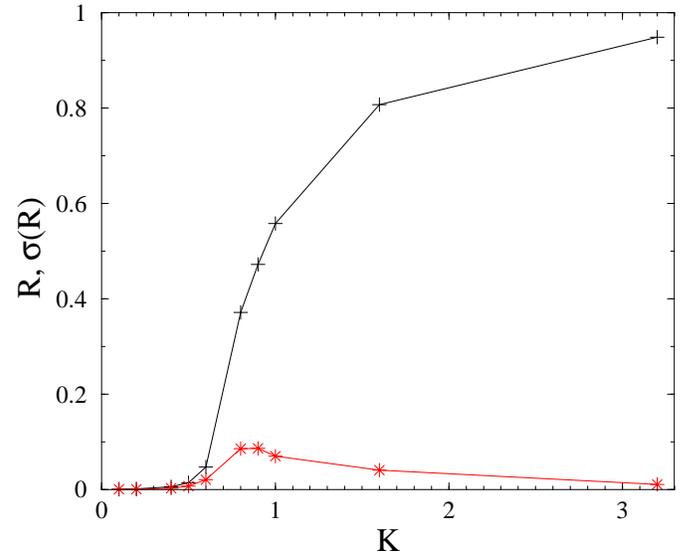}
	\caption{Steady state Kuramoto order parameter and its variance at
        $\alpha=3$ for different $K$ values for $L=100$. Inset: $R(t,L=100)$.}
        \label{beta3D_L100_a3}
\end{figure}

%
\bibliography{bib}

\begin{thebibliography}{29}
\expandafter\ifx\csname natexlab\endcsname\relax\def\natexlab#1{#1}\fi
\expandafter\ifx\csname bibnamefont\endcsname\relax
  \def\bibnamefont#1{#1}\fi
\expandafter\ifx\csname bibfnamefont\endcsname\relax
  \def\bibfnamefont#1{#1}\fi
\expandafter\ifx\csname citenamefont\endcsname\relax
  \def\citenamefont#1{#1}\fi
\expandafter\ifx\csname url\endcsname\relax
  \def\url#1{\texttt{#1}}\fi
\expandafter\ifx\csname urlprefix\endcsname\relax\def\urlprefix{URL }\fi
\providecommand{\bibinfo}[2]{#2}
\providecommand{\eprint}[2][]{\url{#2}}

\bibitem[{\citenamefont{Pikovsky et~al.}(2003)\citenamefont{Pikovsky, Kurths,
  Rosenblum, and Kurths}}]{Pik}
\bibinfo{author}{\bibfnamefont{A.}~\bibnamefont{Pikovsky}},
  \bibinfo{author}{\bibfnamefont{J.}~\bibnamefont{Kurths}},
  \bibinfo{author}{\bibfnamefont{M.}~\bibnamefont{Rosenblum}},
  \bibnamefont{and} \bibinfo{author}{\bibfnamefont{J.}~\bibnamefont{Kurths}},
  \emph{\bibinfo{title}{Synchronization: A Universal Concept in Nonlinear
  Sciences}}, Cambridge Nonlinear Science Series (\bibinfo{publisher}{Cambridge
  University Press}, \bibinfo{year}{2003}), ISBN \bibinfo{isbn}{9780521533522}.

\bibitem[{\citenamefont{Acebr{\'o}n et~al.}(2005)\citenamefont{Acebr{\'o}n,
  Bonilla, Vicente, Ritort, and Spigler}}]{Acebron}
\bibinfo{author}{\bibfnamefont{J.~A.} \bibnamefont{Acebr{\'o}n}},
  \bibinfo{author}{\bibfnamefont{L.~L.} \bibnamefont{Bonilla}},
  \bibinfo{author}{\bibfnamefont{C.~J.~P.} \bibnamefont{Vicente}},
  \bibinfo{author}{\bibfnamefont{F.}~\bibnamefont{Ritort}}, \bibnamefont{and}
  \bibinfo{author}{\bibfnamefont{R.}~\bibnamefont{Spigler}},
  \bibinfo{journal}{Rev. Mod. Phys.} \textbf{\bibinfo{volume}{77}},
  \bibinfo{pages}{137} (\bibinfo{year}{2005}).

\bibitem[{\citenamefont{Arenas et~al.}(2008)\citenamefont{Arenas,
  D{\'\i}az-Guilera, Kurths, Moreno, and Zhou}}]{ARENAS200893}
\bibinfo{author}{\bibfnamefont{A.}~\bibnamefont{Arenas}},
  \bibinfo{author}{\bibfnamefont{A.}~\bibnamefont{D{\'\i}az-Guilera}},
  \bibinfo{author}{\bibfnamefont{J.}~\bibnamefont{Kurths}},
  \bibinfo{author}{\bibfnamefont{Y.}~\bibnamefont{Moreno}}, \bibnamefont{and}
  \bibinfo{author}{\bibfnamefont{C.}~\bibnamefont{Zhou}},
  \bibinfo{journal}{Phys.~Rep.} \textbf{\bibinfo{volume}{469}},
  \bibinfo{pages}{93} (\bibinfo{year}{2008}).

\bibitem[{\citenamefont{Kuramoto}(1975)}]{kuramoto}
\bibinfo{author}{\bibfnamefont{Y.}~\bibnamefont{Kuramoto}}, in
  \emph{\bibinfo{booktitle}{Proceedings of the International Symposium on
  Mathematical Problems in Theoretical Physics}}, edited by
  \bibinfo{editor}{\bibfnamefont{H.}~\bibnamefont{Araki}}
  (\bibinfo{publisher}{Springer, New York}, \bibinfo{year}{1975}).

\bibitem[{\citenamefont{Ott and Antonsen}(2008)}]{oa}
\bibinfo{author}{\bibfnamefont{E.}~\bibnamefont{Ott}} \bibnamefont{and}
  \bibinfo{author}{\bibfnamefont{T.~M.} \bibnamefont{Antonsen}},
  \bibinfo{journal}{CHAOS} \textbf{\bibinfo{volume}{18}},
  \bibinfo{pages}{037113} (\bibinfo{year}{2008}).

\bibitem[{\citenamefont{Hong et~al.}(2007)\citenamefont{Hong, Chat{\'e}, Park,
  and Tang}}]{chate_prl}
\bibinfo{author}{\bibfnamefont{H.}~\bibnamefont{Hong}},
  \bibinfo{author}{\bibfnamefont{H.}~\bibnamefont{Chat{\'e}}},
  \bibinfo{author}{\bibfnamefont{H.}~\bibnamefont{Park}}, \bibnamefont{and}
  \bibinfo{author}{\bibfnamefont{L.-H.} \bibnamefont{Tang}},
  \bibinfo{journal}{Phys.~Rev.~Lett.} \textbf{\bibinfo{volume}{99}},
  \bibinfo{pages}{184101} (\bibinfo{year}{2007}).

\bibitem[{\citenamefont{Choi et~al.}(2013)\citenamefont{Choi, Ha, and
  Kahng}}]{cmk2016}
\bibinfo{author}{\bibfnamefont{C.}~\bibnamefont{Choi}},
  \bibinfo{author}{\bibfnamefont{M.}~\bibnamefont{Ha}}, \bibnamefont{and}
  \bibinfo{author}{\bibfnamefont{B.}~\bibnamefont{Kahng}},
  \bibinfo{journal}{Physical Review E - Statistical, Nonlinear, and Soft Matter
  Physics} \textbf{\bibinfo{volume}{88}} (\bibinfo{year}{2013}).

\bibitem[{\citenamefont{Sakaguchi et~al.}(1987)\citenamefont{Sakaguchi,
  Shinomoto, and Kuramoto}}]{sakaguchi}
\bibinfo{author}{\bibfnamefont{H.}~\bibnamefont{Sakaguchi}},
  \bibinfo{author}{\bibfnamefont{S.}~\bibnamefont{Shinomoto}},
  \bibnamefont{and} \bibinfo{author}{\bibfnamefont{Y.}~\bibnamefont{Kuramoto}},
  \bibinfo{journal}{Prog. Theor. Phys.} \textbf{\bibinfo{volume}{77}},
  \bibinfo{pages}{1005} (\bibinfo{year}{1987}).

\bibitem[{\citenamefont{Hong et~al.}(2005)\citenamefont{Hong, Park, and
  Choi}}]{HPCE}
\bibinfo{author}{\bibfnamefont{H.}~\bibnamefont{Hong}},
  \bibinfo{author}{\bibfnamefont{H.}~\bibnamefont{Park}}, \bibnamefont{and}
  \bibinfo{author}{\bibfnamefont{M.}~\bibnamefont{Choi}},
  \bibinfo{journal}{Phys. Rev. E} \textbf{\bibinfo{volume}{72}},
  \bibinfo{pages}{036217} (\bibinfo{year}{2005}).

\bibitem[{\citenamefont{Juh{\'a}sz et~al.}(2019)\citenamefont{Juh{\'a}sz,
  Kelling, and {\'O}dor}}]{Kurcikk}
\bibinfo{author}{\bibfnamefont{R.}~\bibnamefont{Juh{\'a}sz}},
  \bibinfo{author}{\bibfnamefont{J.}~\bibnamefont{Kelling}}, \bibnamefont{and}
  \bibinfo{author}{\bibfnamefont{G.}~\bibnamefont{{\'O}dor}},
  \bibinfo{journal}{Journal of Statistical Mechanics: Theory and Experiment}
  \textbf{\bibinfo{volume}{2019}}, \bibinfo{pages}{053403}
  (\bibinfo{year}{2019}),
  \urlprefix\url{https://doi.org/10.1088%2F1742-5468%2Fab16c3}.

\bibitem[{\citenamefont{Filatrella et~al.}(2008)\citenamefont{Filatrella,
  Nielsen, and Pedersen}}]{fila}
\bibinfo{author}{\bibfnamefont{G.}~\bibnamefont{Filatrella}},
  \bibinfo{author}{\bibfnamefont{A.~H.} \bibnamefont{Nielsen}},
  \bibnamefont{and} \bibinfo{author}{\bibfnamefont{N.~F.}
  \bibnamefont{Pedersen}}, \bibinfo{journal}{Eur.~Phys.~J.~B}
  \textbf{\bibinfo{volume}{61}}, \bibinfo{pages}{485} (\bibinfo{year}{2008}).

\bibitem[{\citenamefont{Tanaka et~al.}(1997)\citenamefont{Tanaka, Lichtenberg,
  and S.}}]{TL97}
\bibinfo{author}{\bibfnamefont{H.-A.} \bibnamefont{Tanaka}},
  \bibinfo{author}{\bibfnamefont{A.~J.} \bibnamefont{Lichtenberg}},
  \bibnamefont{and} \bibinfo{author}{\bibfnamefont{O.}~\bibnamefont{S.}},
  \bibinfo{journal}{Phys. Rev. Lett.} \textbf{\bibinfo{volume}{78}},
  \bibinfo{pages}{2104–} (\bibinfo{year}{1997}).

\bibitem[{\citenamefont{{\'O}dor and Hartmann}(2018)}]{POWcikk}
\bibinfo{author}{\bibfnamefont{G.}~\bibnamefont{{\'O}dor}} \bibnamefont{and}
  \bibinfo{author}{\bibfnamefont{B.}~\bibnamefont{Hartmann}},
  \bibinfo{journal}{Physical Review E} \textbf{\bibinfo{volume}{98}},
  \bibinfo{pages}{022305} (\bibinfo{year}{2018}).

\bibitem[{\citenamefont{Villegas et~al.}(2014)\citenamefont{Villegas, Moretti,
  and Munoz}}]{Frus}
\bibinfo{author}{\bibfnamefont{P.}~\bibnamefont{Villegas}},
  \bibinfo{author}{\bibfnamefont{P.}~\bibnamefont{Moretti}}, \bibnamefont{and}
  \bibinfo{author}{\bibfnamefont{M.~A.} \bibnamefont{Munoz}},
  \bibinfo{journal}{Sci. Rep.} \textbf{\bibinfo{volume}{4}}, \bibinfo{pages}{1}
  (\bibinfo{year}{2014}).

\bibitem[{\citenamefont{Mill{\'a}n et~al.}(2018)\citenamefont{Mill{\'a}n,
  Torres, and Bianconi}}]{FrusB}
\bibinfo{author}{\bibfnamefont{A.~P.} \bibnamefont{Mill{\'a}n}},
  \bibinfo{author}{\bibfnamefont{J.~J.} \bibnamefont{Torres}},
  \bibnamefont{and} \bibinfo{author}{\bibfnamefont{G.}~\bibnamefont{Bianconi}},
  \bibinfo{journal}{Sci. Rep.} \textbf{\bibinfo{volume}{8}}, \bibinfo{pages}{1}
  (\bibinfo{year}{2018}).

\bibitem[{\citenamefont{\'Odor et~al.}(2022)\citenamefont{\'Odor, Deng,
  Hartmann, and Kelling}}]{USAEUpowcikk}
\bibinfo{author}{\bibfnamefont{G.}~\bibnamefont{\'Odor}},
  \bibinfo{author}{\bibfnamefont{S.}~\bibnamefont{Deng}},
  \bibinfo{author}{\bibfnamefont{B.}~\bibnamefont{Hartmann}}, \bibnamefont{and}
  \bibinfo{author}{\bibfnamefont{J.}~\bibnamefont{Kelling}},
  \bibinfo{journal}{Phys.~Rev.~E} \textbf{\bibinfo{volume}{106}},
  \bibinfo{pages}{034311} (\bibinfo{year}{2022}),
  \urlprefix\url{https://link.aps.org/doi/10.1103/PhysRevE.106.034311}.

\bibitem[{\citenamefont{Mermin and Wagner}(1966)}]{PhysRevLett.17.1133}
\bibinfo{author}{\bibfnamefont{N.~D.} \bibnamefont{Mermin}} \bibnamefont{and}
  \bibinfo{author}{\bibfnamefont{H.}~\bibnamefont{Wagner}},
  \bibinfo{journal}{Phys. Rev. Lett.} \textbf{\bibinfo{volume}{17}},
  \bibinfo{pages}{1133} (\bibinfo{year}{1966}),
  \urlprefix\url{https://link.aps.org/doi/10.1103/PhysRevLett.17.1133}.

\bibitem[{\citenamefont{{\'O}dor and Hartmann}(2020)}]{Powfailcikk}
\bibinfo{author}{\bibfnamefont{G.}~\bibnamefont{{\'O}dor}} \bibnamefont{and}
  \bibinfo{author}{\bibfnamefont{B.}~\bibnamefont{Hartmann}},
  \bibinfo{journal}{Entropy} \textbf{\bibinfo{volume}{22}},
  \bibinfo{pages}{666} (\bibinfo{year}{2020}).

\bibitem[{\citenamefont{Cardy}(1981)}]{Cardy_1981}
\bibinfo{author}{\bibfnamefont{J.~L.} \bibnamefont{Cardy}},
  \bibinfo{journal}{Journal of Physics A: Mathematical and General}
  \textbf{\bibinfo{volume}{14}}, \bibinfo{pages}{1407} (\bibinfo{year}{1981}).

\bibitem[{\citenamefont{Janssen et~al.}(2004)\citenamefont{Janssen, M\"uller,
  and Stenull}}]{PhysRevE.70.026114}
\bibinfo{author}{\bibfnamefont{H.-K.} \bibnamefont{Janssen}},
  \bibinfo{author}{\bibfnamefont{M.}~\bibnamefont{M\"uller}}, \bibnamefont{and}
  \bibinfo{author}{\bibfnamefont{O.}~\bibnamefont{Stenull}},
  \bibinfo{journal}{Physical Review E} \textbf{\bibinfo{volume}{70}},
  \bibinfo{pages}{026114} (\bibinfo{year}{2004}).

\bibitem[{\citenamefont{Chan et~al.}(2015)\citenamefont{Chan, Ghanbarnejad, and
  Grassberger}}]{Gras-av}
\bibinfo{author}{\bibfnamefont{W.}~\bibnamefont{Chan}},
  \bibinfo{author}{\bibfnamefont{F.}~\bibnamefont{Ghanbarnejad}},
  \bibnamefont{and}
  \bibinfo{author}{\bibfnamefont{P.}~\bibnamefont{Grassberger}},
  \bibinfo{journal}{Nature Physics} \textbf{\bibinfo{volume}{11}},
  \bibinfo{pages}{936–} (\bibinfo{year}{2015}).

\bibitem[{\citenamefont{G\'omez-Garde\~nes
  et~al.}(2011)\citenamefont{G\'omez-Garde\~nes, G\'omez, Arenas, and
  Moreno}}]{PhysRevLett.106.128701}
\bibinfo{author}{\bibfnamefont{J.}~\bibnamefont{G\'omez-Garde\~nes}},
  \bibinfo{author}{\bibfnamefont{S.}~\bibnamefont{G\'omez}},
  \bibinfo{author}{\bibfnamefont{A.}~\bibnamefont{Arenas}}, \bibnamefont{and}
  \bibinfo{author}{\bibfnamefont{Y.}~\bibnamefont{Moreno}},
  \bibinfo{journal}{Physical Review Letters} \textbf{\bibinfo{volume}{106}},
  \bibinfo{pages}{128701} (\bibinfo{year}{2011}).

\bibitem[{\citenamefont{Coutinho et~al.}(2013)\citenamefont{Coutinho, Goltsev,
  Dorogovtsev, and Mendes}}]{PhysRevE.87.032106}
\bibinfo{author}{\bibfnamefont{B.~C.} \bibnamefont{Coutinho}},
  \bibinfo{author}{\bibfnamefont{A.~V.} \bibnamefont{Goltsev}},
  \bibinfo{author}{\bibfnamefont{S.~N.} \bibnamefont{Dorogovtsev}},
  \bibnamefont{and} \bibinfo{author}{\bibfnamefont{J.~F.~F.}
  \bibnamefont{Mendes}}, \bibinfo{journal}{Physical Review E}
  \textbf{\bibinfo{volume}{87}}, \bibinfo{pages}{032106}
  (\bibinfo{year}{2013}).

\bibitem[{\citenamefont{{\'O}dor and de~Simoni}(2021)}]{odorSim21}
\bibinfo{author}{\bibfnamefont{G.}~\bibnamefont{{\'O}dor}} \bibnamefont{and}
  \bibinfo{author}{\bibfnamefont{B.}~\bibnamefont{de~Simoni}},
  \bibinfo{journal}{Phys.~Rev.~Res.} \textbf{\bibinfo{volume}{3}},
  \bibinfo{pages}{013106} (\bibinfo{year}{2021}).

\bibitem[{\citenamefont{Grainger and Stevenson}(1994)}]{swing}
\bibinfo{author}{\bibfnamefont{J.}~\bibnamefont{Grainger}, \bibfnamefont{J.}}
  \bibnamefont{and} \bibinfo{author}{\bibfnamefont{D.}~\bibnamefont{Stevenson},
  \bibfnamefont{W.}}, \emph{\bibinfo{title}{Power system analysis}}
  (\bibinfo{publisher}{McGraw-Hill}, \bibinfo{year}{1994}), ISBN
  \bibinfo{isbn}{ISBN 978-0-07-061293-8}.

\bibitem[{\citenamefont{Ahnert and Mulansky}()}]{boostOdeInt}
\bibinfo{author}{\bibfnamefont{K.}~\bibnamefont{Ahnert}} \bibnamefont{and}
  \bibinfo{author}{\bibfnamefont{M.}~\bibnamefont{Mulansky}},
  \emph{\bibinfo{title}{Boost::odeint}}, \urlprefix\url{https://odeint.com}.

\bibitem[{\citenamefont{Jeffrey et~al.}(to be published)\citenamefont{Jeffrey,
  Deng, Barancsuk, Hartmann, and {\'{O}}dor}}]{texpaper}
\bibinfo{author}{\bibfnamefont{K.}~\bibnamefont{Jeffrey}},
  \bibinfo{author}{\bibfnamefont{S.}~\bibnamefont{Deng}},
  \bibinfo{author}{\bibfnamefont{L.}~\bibnamefont{Barancsuk}},
  \bibinfo{author}{\bibfnamefont{B.}~\bibnamefont{Hartmann}}, \bibnamefont{and}
  \bibinfo{author}{\bibfnamefont{G.}~\bibnamefont{{\'{O}}dor}}
  (\bibinfo{year}{to be published}).

\bibitem[{\citenamefont{{\'O}dor et~al.}(2022)\citenamefont{{\'O}dor, Deco, and
  Kelling}}]{Flycikk}
\bibinfo{author}{\bibfnamefont{G.}~\bibnamefont{{\'O}dor}},
  \bibinfo{author}{\bibfnamefont{G.}~\bibnamefont{Deco}}, \bibnamefont{and}
  \bibinfo{author}{\bibfnamefont{J.}~\bibnamefont{Kelling}},
  \bibinfo{journal}{Phys. Rev. Research} \textbf{\bibinfo{volume}{4}},
  \bibinfo{pages}{023057} (\bibinfo{year}{2022}).

\bibitem[{\citenamefont{{\'O}dor et~al.}(2021)\citenamefont{{\'O}dor, Kelling,
  and Deco}}]{KKI18deco}
\bibinfo{author}{\bibfnamefont{G.}~\bibnamefont{{\'O}dor}},
  \bibinfo{author}{\bibfnamefont{J.}~\bibnamefont{Kelling}}, \bibnamefont{and}
  \bibinfo{author}{\bibfnamefont{G.}~\bibnamefont{Deco}},
  \bibinfo{journal}{J.~Neurocomputing} \textbf{\bibinfo{volume}{461}},
  \bibinfo{pages}{696} (\bibinfo{year}{2021}).

\end{thebibliography}

\end{document}